\documentstyle[12pt,fleqn]{article}
\begin{document}
\baselineskip= 22 truept
\def\be{\begin{equation}}
\def\ee{\end{equation}}
\def\bea{\begin{eqnarray}}
\def\eea{\end{eqnarray}}
\def\pa{\partial}
\begin{titlepage}
\begin{flushright}
IP/BBSR/95-18 \\
hep-th/9503154 \\
\end{flushright}
\vspace{1cm}\begin{center} {\large \bf Ehlers Transformations
and String Effective Action}\\ \vspace{1cm} {\bf Alok Kumar and
Koushik Ray }\\ Institute of Physics, \\Bhubaneswar 751 005,
INDIA \\ email: kumar, koushik@iopb.ernet.in \\ \today
\end{center}
\thispagestyle{empty}
\vskip 4cm
\begin{abstract}

We explicitly obtain the generalization of the Ehlers
transformation for stationary axisymmetric Einstein
equations to string theory.  This is accomplished by finding the
twist potential corresponding to the moduli fields in the
effective two dimensional theory.  Twist potential and symmetric
moduli are shown to transform under an $O(d,d)$ which is a
manifest symmetry of the equations of motion. The non-trivial
action of this $O(d,d)$ is given by the Ehlers transformation
and belongs to the set $O(d) \times O(d)\over O(d) $.
\end{abstract}
\vfil
\end{titlepage}
\eject

Symmetries of stationary axisymmetric Einstein equations
have been extensively studied in past decades\cite{xan,ernst}.
An interesting aspect of this system is the existence of an
infinite dimensional symmetry, known as the Geroch
group\cite{geroch}. It has also been realized that the infinite
symmetry structure in this case renders the
model integrable\cite{hn}.
This system describes an Ernst Sigma model\cite{xan,ernst}
which has been
studied a great deal including for the supergravity models in
two dimensions. The existence of an infinite number of conserved
currents and their algebraic structure have also been
analyzed\cite{hn,nicolai}.  These symmetry transformations have
also been used for generating solutions of Einstein equations
such as Tomimatsu-Sato class of metrics\cite{ste}.

Some aspects of the Ernst sigma model and the infinite
dimensional symmetries have been analyzed recently in the
context of string theory\cite{bakas,pui,jmu,sen}.
In particular, the
duality symmetries of the four dimesnsional string theory have
been shown to be a subgroup of the infinite dimensional String
Geroch group in the presence of two commuting
isometries.

In this paper we further study the symmetries of string theory
and obtain the analogue of the Ehlers transformation. We also
show that, as in the case of Einstein equations, this Ehlers
transformation is the one responsible for the existence of an
infinite number of conserved currents in the theory and for
generating new solutions. We in fact show the existence of a new
$O(d,d)$ group of symmetry transformations of which Ehlers is a
subset. The present approach has the advantage
that we can write
down the finite transformations of the fields directly.

An interesting scheme for finding out new solutions of stationary
axisymmetric Einstein equations through the underlying symmetry
transformations was developed in a series of papers by
Kinnersley et al\cite{kinn,kc}. They showed the existence of two
sets of $SL(2,R)$ symmetries in this theory denoted as the
groups $\cal G$ and $\cal H$. The action of two of the
generators of $\cal G$
and $\cal H$, which correspond to translation and scaling of the
corresponding variables, coincide. However, the third generators
of these groups do not commute with the dualization, {\it viz.} the
Kramer-Neugebauer(KN) mapping\cite{kn}. This fact gives rise to
an infinite set of conserved currents.

In this paper we generalize the approach of \cite{kinn,kc} to
string theory. We find out a twist potential corresponding to
the antisymmetric moduli field. It is also shown that the
equations of motion, written in terms of the twist potential,
are similar to the original ones and are in fact related by a
generalization of the KN-mapping. We show the existence of two
sets of $O(d, d)$ symmetries. The first one acts on the usual
moduli of string theory. The second $O(d, d)$ acts on the moduli
involving the twist potentials.  These are the analogues of the
groups $\cal G$ and $\cal H$ in the present case. Now the set of
generators corresponding to the constant shift of the
antisymmetric tensor and the constant coordinate transformation
in $\cal G$ and $\cal H$ coincide. The action of the remaining
one in $\cal H$
is identified as the generalized Ehlers transformation. We also
write down the action of the generalized Ehlers transformation
on the original field variables.

We now begin our study of the bosonic part of the four
dimensional heterotic string effective action for the case when
the $E_8 \times E_8$ gauge field backgrounds are set to zero. In
the presence of two commuting isometries, the equations of
motion of this theory coincide with those following from the two
dimensional string effective action. Therefore, we concentrate
on the 2D action and the corresponding equations of motion. We
further restrict to the case when the two dimensional gauge
fields have vanishing background.  This is consistent with the
fact that gauge fields have no physical degrees of freedom in
two dimensions. The action can then be written as
\begin{equation} \label{action}
S = \int d^2 x \sqrt{g} e^{-2\phi} \left[ R + 4 g^{\mu \nu}
\partial_{\mu}\phi \partial_\nu\phi + \frac{1}{8} g^{\mu \nu}
Tr( \partial_\mu M^{-1} \partial_\nu M) \right],
\end{equation}
where the matrix $M$, representing the moduli $G$ and $B$, is
parametrized as
\begin{equation}
M = \left( \begin{array}{cc} G^{-1} & -G^{-1} B \\ B G ^{-1} & G
-  B G^{-1} B \end{array} \right).
\end{equation}
$G$ and $B$ are respectively $d\times d$ symmetric and
antisymmetric matrix-valued scalar fields. For the heterotic
string $d = 8$, but we keep it arbitrary in the present
discussion.
The equations of motion for the above action can be written
as
\begin{eqnarray}
\label{eom1}
\partial_{\mu} ( \sqrt{g} g^{\mu \nu }
e^{-2\phi} M^{-1} \partial_\nu
M ) = 0 \\ \label{eom2}
\partial_\mu ( \sqrt{g} g^{\mu \nu} \partial_\nu e^{-2\phi} ) =
0 \\
\label{eom3}
R^{(2)}_{\mu \nu} + 2\nabla_\mu \nabla_\nu \phi + \frac{1}{8} Tr
(\partial_\mu M^{-1} \partial_\nu M) = 0.
\end{eqnarray}

The form of these equations are similar to the ones for the
Einstein equations in \cite{kinn} and the matrix $M$ plays the
role of the $2\times 2$ metric in the case of Einstein
equations. The equations (\ref{eom1}) and (\ref{eom2}) can be
rewritten in the conformal gauge, with an identification
$ \rho = e^{-2\phi}$, as
\begin{eqnarray}
\label{eom11}
\partial^{\mu} ( \rho M^{-1} \partial_\mu M ) = 0, \\
\label{eom22} \partial^\mu \partial_\mu \rho = 0.
\end{eqnarray}
The third equation (\ref{eom3}) turns out to be an integrable
one for the conformal factor and takes the following form in the
conformal gauge $ g_{\mu \nu } = e^{2\Gamma} \delta_{\mu \nu} $:
\begin{eqnarray} \label{eom33}
- \delta_{\mu \nu} \partial^\sigma \partial_\sigma \Gamma + 2
(\partial_\mu \partial_\nu \phi &-& \partial_\mu \phi \partial_\nu
\Gamma - \partial_\nu \phi \partial_\mu \Gamma + \delta_{\mu \nu}
\partial^\sigma \Gamma \partial_\sigma \phi) \nonumber \\
&+& \frac{1}{8} Tr ( \partial_\mu M^{-1} \partial_\nu M) = 0.
\end{eqnarray}
Determination of this conformal factor is important for the
generation of new solutions. Here we have chosen the two
dimensional metric to be Euclidean. This corresponds to the
contraction of the Levi-Civita tensor:
$\epsilon^{\mu \alpha}\epsilon_{\mu \beta}=
\delta^{\alpha}_{\beta}$. All the results go through for
the Minkowski space as well with few changes in signs.
We now concentrate on equations
(\ref{eom11}), (\ref{eom22}) and will come back to
(\ref{eom33}) later.

Equation (\ref{eom11}) and (\ref{eom22}) are the same as that for
the Ernst sigma
model\cite{xan,nicolai,bakas}. It is also clear that these
equations are invariant under
an $O(d, d)$ group of symmetry transformations:
\begin{equation}
M \rightarrow \tilde{M} = \Omega M \Omega^T,
\end{equation}
which includes the T-duality symmetry of the string theory as its
discrete subgroup $O(d, d;Z)$ and has been used for generating
solutions. The non-trivial action of T-duality is represented
by a group $O(d;Z) \times O(d;Z) \over O(d;Z)$\cite{asen}. The
T-duality is
also conjectured to be a symmetry, not only of the string
effective action, but of the full string theory\cite{sch} too.

The Ehlers transformation on the moduli is now determined on the
lines of \cite{kinn} by obtaining the twist potentials. For this we
observe that equations (\ref{eom11}) are total divergence
conditions and when written in terms of $G$ and $B$, take the
form
\begin{eqnarray} \label{gb1}
\partial^\mu ( \rho G^{-1} \partial_\mu B G^{-1}) = 0 \\
\label{gb2} \partial^\mu[ \rho (G \partial_\mu G^{-1} + B G^{-1}
\partial_\mu B G^{-1})] = 0 \\ \label{gb3} \partial^\mu[ \rho(
\partial_\mu B + B G^{-1} \partial_\mu B G^{-1} B + G
\partial_\mu G^{-1} B - B G^{-1} \partial_\mu G) ] = 0 .
\end{eqnarray}
One can also show that there are only two algebraically
independent equations among (\ref{gb1})--(\ref{gb3}), which is
due to the fact that the $M$-equation is essentially a combination
of the two equations for $G$ and $B$.  One can now define a set
of three potentials by the following relations:
\begin{eqnarray} \label{si1}
\rho G^{-1} \partial_\mu B G^{-1} = - \epsilon_{\mu \nu }
\partial^\nu \psi_1, \\ \label{si2} \rho (G \partial_\mu G^{-1} +
B G^{-1} \partial_\mu B G^{-1}) = - \epsilon_{\mu \nu}
\partial^\nu \psi_2, \\ \label{si3} \rho (\partial_\mu B + B
G^{-1} \partial_\mu B G^{-1} B + G \partial_\mu G^{-1} B - B
G^{-1} \partial_\mu G ) = - \epsilon_{\mu \nu} \partial^\nu
\psi_3,
\end{eqnarray}
such that the equations (\ref{gb1})--(\ref{gb3}) are the
integrability conditions for the $\psi_i$'s. Equations
(\ref{si1})--(\ref{si3}) are matrix generalization of the
dualization relations in \cite{kinn} and the {\it twist
potential} $\psi\equiv\psi_1$ is the dual field corresponding to
the antisymmetric moduli $B$.  One can use $G$, $\psi$ and
$\rho$ to rewrite the equations (\ref{gb1})--(\ref{gb3}). For
example, an equation similar to (\ref{gb1}) can be obtained by
inverting (\ref{si1}) as,
\begin{equation} \label{gsi1}
\partial^\mu (\frac{G}{\rho} \partial_\mu \psi G ) = 0.
\end{equation}
Equation (\ref{gb2}), when expressed in terms of $\psi$, takes
the form
\begin{equation} \label{gsi2}
\partial^{\mu} (\rho G \partial_{\mu} G^{-1} ) + \frac{1}{\rho} G
\partial_\mu\psi G \partial^\mu \psi = 0.
\end{equation}
Then using (\ref{gsi1}), (\ref{gsi2}) and (\ref{eom22}), the
last equation, (\ref{gb3}), is satisfied identically.

The equations of motion in $G, \psi$ variables, (\ref{gsi1}) and
(\ref{gsi2}), can also be expressed as
\begin{equation} \label{main}
\partial^\mu(\rho M'^{-1} \partial_\mu M') = 0,
\end{equation}
where
\begin{equation} \label{mprime}
M' = \left(
\begin{array}{cc}
\frac{G}{\rho}  & \frac{G}{\rho} \psi \\
-\psi \frac{G}{\rho} & - \rho G^{-1} -
\psi \frac{G}{\rho} \psi \end{array} \right).
\end{equation}
Equation (\ref{main}) is one of the main results of this paper
as it leads to new symmetries of this theory. The
equations in terms of $M'$ are
related to those in terms of
$M$ through a KN-mapping of the form
$G \rightarrow \rho G^{-1}, \quad  B \rightarrow i\psi$.

One can also introduce a set of three potentials for $M'$ as
\begin{eqnarray} \label{bsi1}
 \epsilon_{\mu \nu} \partial^\nu \Psi_1 &=& - \frac{1}{\rho} G
\partial_{\mu}\psi G,
\\ \label{bsi2}  \epsilon_{\mu \nu} \partial^\nu \Psi_2 &=&
- G \partial_\mu (\rho G^{-1}) - \frac{G}{\rho} \pa_\mu \psi G
\psi,\\ \label{bsi3} \epsilon_{\mu \nu} \partial^\nu
\Psi_3 = -\rho^2 G^{-1}\partial_{\mu}(\frac{G}{\rho})\psi &-& \rho
\partial_{\mu}\psi + \psi G \partial_{\mu}{(\rho G^{-1})} + \psi G
(\partial_{\mu}\psi) {G\over \rho}\psi,
\end{eqnarray}
whose integrability conditions give back the equations of motions
for $G$ and $\psi$.
Comapring (\ref{bsi1}) with (\ref{si1}) we get $\Psi_1 = B$.
Equation (\ref{eom3}) can also be rewritten in terms of $G$
and $\psi$. For this we first rewrite (\ref{eom33}) using
(\ref{gsi2}) as
\begin{eqnarray} \label{ita}
- \delta_{\mu \nu} \partial^\sigma \partial_\sigma \Gamma &+& 2
\left( \partial_\mu \partial_\nu \phi - \partial_\mu \phi
\partial_\nu  \Gamma - \partial_\nu \phi \partial_\mu \Gamma +
\delta_{\mu \nu} \partial^\sigma \Gamma \partial_\sigma \phi
\right) \nonumber \\ & + & \frac{1}{4 \rho^2} \delta_{\mu \nu} Tr
\left( \partial^\sigma \psi G \partial_\sigma \psi G \right)
\nonumber \\ & - & \frac{1}{4} Tr \left( G^{-1} \partial_\mu G
G^{-1} \partial_\nu G + \frac{1}{\rho^2} \partial_\mu \psi G
\partial_\nu \psi G \right) = 0.
\end{eqnarray}
It can be shown that under a transformation
\begin{equation}
\Gamma' = \Gamma - \frac{1}{4} \ln \det G  - \frac{d}{4} \phi,
\end{equation}
equation (\ref{ita}) goes over to
\begin{eqnarray}
- \delta_{\mu \nu} \partial^\sigma \partial_\sigma \Gamma' &+& 2
\left[ \partial_\mu \partial_\nu \phi - \partial_\mu \phi
\partial_\nu \Gamma' - \partial_\nu \phi \partial_\mu \Gamma' +
\delta_{\mu \nu} \partial^\sigma \Gamma' \partial_\sigma \phi
\right] \nonumber \\ & - & \frac{1}{2} \left[ \partial_\mu
\phi \partial_\nu (\ln \det G ) + \partial_\nu \phi
\partial_\mu ( \ln \det G ) + 2d \partial_\mu \phi \partial_\nu
\phi \right] \nonumber \\ & - & \frac{1}{4} Tr \left( G^{-1}
\partial_\mu G G^{-1} \partial_\nu G + \frac{1}{\rho^2}
\partial_\mu \psi G \partial_\nu \psi G\right) = 0.
\end{eqnarray}
which can be further rewritten as
\begin{equation} \label{gsi3}
R_{\mu \nu} + 2\nabla_\mu \nabla_\nu \phi + \frac{1}{8}
Tr(\partial_\mu M'^{-1} \partial_\nu M') = 0
\end{equation}
with a new conformal factor $\Gamma'$. (\ref{eom22}),
(\ref{main}) and (\ref{gsi3}) are the complete set of equations
of motion rewritten in terms of $G$ and $\psi$.

We now study the symmetries of equations (\ref{main}). Like the
$M$ equations these are also manifestly invariant under a
transformation
\begin{equation} \label{transf}
M' \rightarrow \tilde{M'} = \Omega' M' \Omega'^T,
\end{equation}
However, since $\psi$ is an antisymmetric matrix due to the
dualization relation (\ref{si1}), $M'$ satisfies
\begin{equation} \label{rest}
        M'LM' = - L, \quad {\rm where} \quad
 L = \left( \begin{array}{cc} 0 & I \\
                         I & 0
\end{array}     \right)
\end{equation}
and $I$ is the $d$ dimensional identity matrix. The condition
(\ref{rest}) is maintained under (\ref{transf}) provided
$\Omega'$ satisfies the relation
\begin{equation}
\Omega'^T L \Omega' = L.
\end{equation}
As a result, the matrices $\Omega'$ belong to the group $O(d,
d)$.

We have therefore shown that a new type of $O(d, d)$ symmetry
transformation can be used for generating solutions of the
string effective action. Like the $o(d, d)$ algebra of the
symmetry transformations in equations (\ref{eom11}), this
is also expected to be the part of an affine $\hat{o}(d,d)$.
We have been able to explicitly identify the action of the new
$O(d, d)$ by making use of dualization.

We now show that only a part of the new $O(d, d)$ has non-trivial
action to give infinite number of conservation laws and generate
new solutions. For this, we classify the independent $O(d, d)$
transformations into three sets:
\begin{equation}
\Omega_1 = \left(\begin{array}{cc} I & \gamma \\
                          0 & I \\
\end{array}\right)  \quad
\Omega_2 = \left(\begin{array}{cc} I & 0 \\
                          \alpha & I \\
\end{array}\right) \quad
\Omega_3 = \left(\begin{array}{cc}  {A^{-1}}^T & 0 \\
					0 & A \\
\end{array}\right)
\end{equation}
where $\gamma$ and $\alpha$ are real
antisymmetric matrices and $A$ is an arbitrary nonsingular real
$d\times d$ matrix.

A different realization of $O(d,d)$ matrices, to study T-duality
symmetries, was given in \cite{asen}, where the non-trivial part
of $O(d,d)$ was identified as $O(d)\times O(d) \over O(d)$. We
have chosen the above representaton in order to make connection
with the Ehlers transformation of General Relativity\cite{kinn}. In
\cite{asen} it was shown that the matrices $\Omega_2$ and
$\Omega_3$ together parametrize the generators of $O(d,d)$
which are outside a set $O(d)\times O(d) \over O(d) $. As a
result, one can identify $\Omega_1 $ as one which
parametrizes $O(d)\times O(d) \over O(d) $. The $O(d, d)$
transformations of $G$ and $\psi$ under the $\Omega_i$'s can be
obtained from the condition $\tilde{M'}_i = \Omega_i M'
\Omega_i^T$. The transformation under $\Omega_1$ is written as
\begin{eqnarray}
{\rm (i)} \quad \frac{G}{\rho} & \rightarrow & \frac{G}{\rho} -
(\frac{G}{\rho} \psi \gamma + \gamma \psi \frac{G}{\rho} ) +
\gamma (\rho G^{-1} + \psi \frac{G}{\rho} \psi ) \gamma
\nonumber \\ - \frac{\psi G }{\rho} & \rightarrow &- \frac{\psi
G}{\rho} + (\rho G^{-1} + \psi \frac{G }{\rho} \psi ) \gamma
\end{eqnarray}
leaving $\rho$, as well as $ (\rho G^{-1} + \psi \frac{G}{\rho}
\psi )$, invariant. The antisymmetry of $\psi$ is ensured as the
condition (\ref{rest}) is preserved under these transformations.
Under $\Omega_2$ and $\Omega_3$ we have, respectively
\begin{eqnarray} \label{trans11}
{\rm (ii)} \quad \frac{G}{\rho} \rightarrow \frac{G}{\rho},
\qquad \psi \rightarrow \psi - \alpha \\ \label{trans12} {\rm
(iii)} \quad \frac{G}{\rho} \rightarrow {A^{-1}}^T
\frac{G}{\rho} A^{-1}, \qquad \psi \rightarrow A \psi A^T.
\end{eqnarray}
The action of $\Omega$'s on $B$-fields is obtained from a
relation $\partial_\mu \tilde{\chi_i} = {\Omega_i}^{{-1}^T}
\partial_\mu \chi
\Omega_i^T$, where
\begin{equation}
\chi = \left( \begin{array}{cc} -\Psi_2^T & \Psi_3 \\ \Psi_1 &
\Psi_2 \end{array} \right).
\end{equation}
Consequently $B \equiv \Psi_1$ has the following transformations
under $\Omega_i$'s.
\begin{eqnarray} \label{trans21}
{\rm (i)} \quad B  \rightarrow \tilde{B} = B - \Psi_2 \gamma
- \gamma \Psi_2^T - \gamma \Psi_3 \gamma, \\ \label{trans22} {\rm
(ii)} \quad B  \rightarrow B   \qquad {\rm and} \qquad
{\rm (iii) } \quad B  \rightarrow  {A^{-1}}^T B A^{-1}.
\end{eqnarray}
{}From (\ref{trans11}) and (\ref{trans22}) we see that $\Omega_2$
simply corresponds to a constant shift in $B$ and $\psi$. The
action of $\Omega_3$ is same as a constant coordinate
transformation. These, together with the results of \cite{asen},
imply that the corresponding parts of $O(d,d)$ acting on $M$
and $M'$ in fact coincide. However, $\Omega_1$ acts very
differently on $B$ and on $\psi$. Its non-local action on $B$
originates from the fact that the potentials $\psi_i$ are related to
$\Psi_i$ through integrations.

The infinite set of conserved currents are now obtained by
applying the transformations $\Omega_i$'s on the dualization
equation (\ref{si1}). It is first observed that the action of
$\Omega_2$ and $\Omega_3$ leaves this equation invariant and
therefore does not lead to any new conservation law. The
transformation by $\Omega_1$ on the other hand implies
\begin{eqnarray}
\left[ \frac{G}{\rho} \right. & - & \left. \left( \frac{G}{\rho}
\psi \gamma + \gamma \psi \frac{G}{\rho} \right) + \gamma \left(
\rho G^{-1} + \psi \frac{G}{\rho} \psi \right) \gamma
\right]^{-1} \partial^\mu \left[ B - \Psi_2 \gamma - \right.
\nonumber \\ \gamma \Psi_2^T & - & \left. \gamma \Psi_3 \gamma
\right] \left[ \frac{G}{\rho} - \left( \frac{G}{\rho} \psi
\gamma + \gamma \psi \frac{G}{\rho} \right) + \gamma \left( \rho
G^{-1} + \psi \frac{G}{\rho} \psi \right) \gamma \right]^{-1}
\nonumber \\ & = & - \epsilon^{\mu \nu} \rho \partial_\nu \left(
\left[ \psi \frac{G}{\rho} - \left( \rho G^{-1} + \psi
\frac{G}{\rho} \psi \right) \gamma \right] \left[ \frac{G}{\rho}
- \left( \frac{G}{\rho} \psi \gamma \right. \right. \right. +
\nonumber \\ & \mbox{} & \qquad \left. \left. \gamma \left.
\psi \frac{G}{\rho} \right) + \gamma \left( \rho G^{-1} + \psi
\frac{G}{\rho} \psi \right) \gamma \right]^{-1} \right).
\end{eqnarray}
The infinite set of conserved currents for this theory is now
obtained by expanding both sides of this equation in powers of
$\gamma$. For example, the zeroth power gives back the
dualization equation
(\ref{si1}) and the conservation equation (\ref{gb1}). From the
next order we get, after some calculations,
\begin{equation}
\epsilon^{\mu \nu}\partial_\nu G = \frac{G}{\rho} (\partial^\mu
\Psi_2 - \partial^\mu B \psi ),
\end{equation}
which leads to another conservation equation:
\begin{equation}
\partial_\mu \left[ \frac{G}{\rho} (\partial^\mu \Psi_2 -
\partial^\mu B \psi ) \right] = 0.
\end{equation}
as well as its transpose.

To conclude, in this paper we have obtained the explicit action
of a new $O(d,d)$ transformation in the string effective action.
We have identified the non-trivial part of this $O(d,d)$ as a
generalized Ehlers transformation. We have also shown that the
generalized Ehlers belongs to a set of $O(d)\times O(d) \over
O(d)$ transformations which is similar to the non-trivial part of
the T-duality. Our results indicate that only a subset of the
infinite dimensional symmetries represented by
$\hat{o}(d,d)$ algebra of these theories has a role in
generating new solutions.

We can also introduce an {\it Ernst} potential, $ E = \rho G^{-1} -
i \psi $, and combine the
equations (\ref{gsi1}) and (\ref{gsi2}) for $G$ and $\psi$ into
a single equation
\begin{equation}
\partial^\mu \left( \rho \partial_\mu E \right) = 2 \rho
\partial^\mu E \left( E  + E^T \right)^{-1} \partial_\mu E.
\end{equation}
This is a matrix generalization of a similar equation for
Einstein gravity\cite{xan}. Hopefully this will lead to an
{\it Ernst formulation} of the present problem.

As mentioned earlier, the results of this paper are also valid
for the choice of the two dimensional metric being
Minkowski. In this case the sign of the first term in
the (22) element of $M'$ in equation (\ref{mprime})
is positive. However, the covariant
forms of all the original as well as the dual equations
of motion remain unchanged.

It will also be interesting to apply the Ehlers transformation
obtained in this paper to various physical situations alone or
together with T-duality and extend the results to heterotic
string theories for non-zero $E_8 \times E_8$ gauge
backgrounds.  In this paper we have related the
equations of motion  of the original and dual
fields. Possibly one can also write down the action and the
equations of motion where both $B$ and $\psi$ appear on an equal
footing. Such actions for the world-sheet theories as well as
for four dimensional theories  have already
been written\cite{sch}. Some of these will
be reported in future.

\section*{Acknowledgement} We thank J. Maharana for invaluable
discussions.

\end{document}